\documentclass[aps,showpacs,superscriptaddress,longbibliography]{revtex4-1}
\usepackage{amsfonts}
\usepackage{amssymb}
\usepackage{amsmath}
\usepackage{graphicx,color}
\usepackage{bm}
\usepackage{ulem}

\begin{document}

\title{ Contribution of Subthreshold States to the Residual Energy Distribution of $^{159}$Dy}

\date{\today}

\author{Tadafumi Kishimoto}
\affiliation{Research Center for Nuclear Physics, Osaka University, Ibaraki, Osaka
567-0047, Japan}

\author{Toru Sato}
\affiliation{Research Center for Nuclear Physics, Osaka University, Ibaraki, Osaka
567-0047, Japan}

\begin{abstract}

We investigate the residual energy distribution in the electron capture
(EC) process of $^{159}$Dy, emphasizing the role of subthreshold atomic states,
which have typically been omitted in conventional spectral analyses. By
incorporating these energetically forbidden hole states into the spectral
function $P(E)$, we demonstrate a significant enhancement - over an order of
magnitude - in the EC rate near the zero-momentum neutrino emission region.
This enhancement increases further with larger $Q$ values, contrary to
standard expectations. Our analysis shows that $P(E)$ can be experimentally
determined, resolving ambiguities near the endpoint and enabling new
approaches to \lq ultra-low $Q$ value\rq \  EC reactions for neutrino mass studies.

\end{abstract}

\maketitle

Observations of neutrino fluxes led to the discovery of neutrino oscillations
\cite{Cleveland_1998,PhysRevLett.81.1562,PhysRevLett.89.011301},   
demonstrating that neutrinos are massive particles. Despite numerous efforts to measure 
the neutrino mass, only upper limits have been established thus far. 
Cosmological studies place an upper limit of approximately 1 eV on the sum of all neutrino masses
\cite{Bennett_2013,ade2016planck}.                 
Similarly, the absence of neutrinoless double beta decay provides an upper limit of 0.04-0.3 eV 
for the Majorana mass
\cite{PhysRevLett.130.051801,PhysRevLett.131.142501}.        
Recent direct measurements have established an upper limit of 0.9 eV for the electron neutrino 
mass \cite{katrin2022direct}.  
Although the mass differences between the three neutrino species are known, the absolute masses 
of the neutrinos remain undetermined.  Measurement of the decay kinematics provides a 
model-independent method for determining the neutrino mass.

Among the three neutrino species, the electron neutrino offers the best opportunity for mass measurement, 
owing to its smallest decay kinetic energy. The influence of the electron antineutrino mass can be observed 
in the shape of the spectrum of beta decay or EC.  
$^3$H decay has been the focus of extensive study due to its $Q$ value of 18.7 keV \cite{katrin2022direct}. 
Even with this low $Q$ value, the fraction of the decay rate in the region of interest (ROI) 
for the energy width ($\Delta E$) at the endpoint region is $({\Delta E\over Q})^3 \sim 10^{-13}$ 
for $\Delta E \sim 1$ eV, where the sub-eV neutrino mass effect is expected to manifest.

$^{187}$Re decay has an even smaller $Q$ value of 2.49 keV \cite{PhysRevC.90.042501}. 
The fraction of ROI $10^{-10}$ is much higher than that of $^3$H. However, its total decay rate 
(1/lifetime) is 9 orders of magnitude smaller than that of $^3$H because the smaller $Q$ value results 
in a smaller total decay rate, and $^3$H undergoes an allowed transition whereas $^{187}$Re undergoes 
a forbidden transition.  The suitability of a particular decay process for study primarily depends on the 
$Q$ value, although other factors must also be considered. 

For EC, the situation differs somewhat. The decay rate of the EC process is proportional 
to $Q^2$, which reflects only the phase space of the emitted neutrino. Consequently, 
a small $Q$ value is more advantageous for EC than for beta decay.  For instance, 
in $^{163}$Ho decay, which has a  $Q$ value of 2.6 keV~\cite{Velte:2019jvx}, 
this characteristic proves advantageous for neutrino mass studies.

Recently, candidates for \lq ultra-low $Q$ value\rq \ beta decay and EC have been 
reported~\cite{redshaw2023precise, PhysRevC.107.015504}. Among these, EC to $^{159}$Dy, 
specifically to the $5/2^-$ excited state of $^{159}$Tb, has been proposed to have the smallest $Q$ 
value~\cite{Ge:2021gnp}.  It is $Q - \Delta_{\rm nucl} = 1.18(19)$ keV, where
$Q=364.93(19)$keV and
$\Delta_{\rm nucl}=363.5449(14)$keV is the energy of the nuclear excited state.

The measurement of the residual energy left in the daughter atom provides
the total energy of the neutrino, enabling the extraction of the neutrino mass from the 
spectrum shape near the endpoint region.  
For EC, however, the spectrum shape is controlled not only 
by the phase space factor of the neutrino but also by the strength distribution of electron hole states.
In the EC process, electrons are absorbed from the $s_{1/2}$ or $p_{1/2}$ 
orbitals, and the residual energy is predominantly determined by the energy of the 
atomic hole state.
For the $^{159}$Dy EC, the binding energy of the atomic N orbit is $0.4$ keV, 
which is only $0.78$ keV away from the point of minimum neutrino energy.

To date, predictions of the neutrino energy spectrum have included only energetically allowed 
hole states ("continuum states"), where the excitation energy of the residual atom $E_{\rm ex}$ 
satisfies $E_{\rm ex} < Q - \Delta_{\rm nucl} - m_\nu$.  In this paper, we highlight the importance 
of including EC to energetically forbidden excited states ("subthreshold states") of the daughter atom. 
We demonstrate that the contribution of these states enhances the partial capture rate by 
more than an order of magnitude near the minimum neutrino energy for $^{159}$Dy.  
This analysis opens new avenues for exploring \lq ultra-low $Q$ value\rq \  EC reactions 
in neutrino mass searches.

The transition matrix element of EC is given as:
\begin{eqnarray}
  T_{{\cal F}{\cal I}} & = & \frac{G_\beta}{\sqrt{2}} \int d\bm{x} <{\cal F}| J_\mu(\bm{x})\bar{u}(\bm{p}_\nu)
    e^{-i\bm{p}_\nu\cdot\bm{x}}\gamma^\mu (1 - \gamma_5)\psi_e(\bm{x})|{\cal I}>.
\end{eqnarray}
Here, $|{\cal I}>$ and ${\cal F}>$ represent the initial and final nuclear and atomic states, respectively.
 $J_\mu$ and $\psi_e(\bm{x})$ are the hadronic weak current density and the electron field operator. 
The rate of EC is given as:
\begin{eqnarray}
  \lambda_{\rm EC} &= & 2\pi \int \frac{d\bm{p}_\nu}{(2\pi)^3}
  \sum_{{\cal F},\bar{{\cal I}}}  \delta(E_{\cal I} - E_{\cal F} - E_\nu)|T_{{\cal F}{\cal I}}|^2,
\end{eqnarray}
where $\sum_{{\cal F},\bar{\cal I}}$ represents the sum over the final states and the average 
over the initial spin states.  The neutrino energy is given as
 $E_\nu = E_{\cal I} - E_{\cal F} = Q - \Delta_{\rm nucl} - E_{\rm ex}$.
The energy differential rate of EC, corresponding to the experimentally obtained energy spectrum, 
is expressed as the product of the nuclear reduced matrix element $C$, the atomic spectral function 
$P(E_{\rm ex})$, and the phase space factor of the neutrino $p_\nu E_\nu$ while neglecting the small 
width of the nuclear excited state:
\begin{eqnarray}
 \frac{d\lambda_{\rm EC}}{dE_\nu} & = & \frac{G_\beta^2}{2\pi}
     p_\nu E_\nu C  P(E_{\rm ex}), \label{eq:dlam}
\end{eqnarray}
where  its integration over neutrino energy gives the EC decay rate: 
\begin{eqnarray}
  \lambda_{\rm EC} & = & \int_{m_\nu}^{Q - \Delta_{\rm nucl}}  \
  \frac{d\lambda_{\rm EC}}{dE_\nu} dE_\nu.
\end{eqnarray}
The response of the atomic system is accounted for through the spectral function,
which is defined by summing over all possible atomic final states with excitation energy $E_{\rm ex}$:
\begin{eqnarray}
  P(E_{\rm ex}) = \sum_{\bar{i}}<i |\psi^\dagger_{e,\alpha}(0) \delta(E_{\rm ex} - H_{\rm atom})
                      \psi_{e,\beta}(0)|i>(1-\gamma_5)_{\alpha,\beta}. \label{eq:spect}
\end{eqnarray}
Here, $|i>$ denotes the atomic ground state of the parent atom,
and $H_{\rm atom}$ is the Hamiltonian of the
daughter atomic system, with the ground state energy set to zero.
$\alpha$ and $\beta$ represent the Dirac components of the
  electron field operator $\psi_e$.
The Dirac matrix $1-\gamma_5$  projects left-handed electrons.
For the allowed Gamow-Teller transition, the dependence on neutrino energy
for the nuclear transition  matrix element can be safely neglected
at low energy. In a leading-order approximation, the lepton wave functions
are evaluated at the origin~\cite{horiuchi2021electron} leading to Eq.~(\ref{eq:spect}).
The nuclear reduced matrix element $C$ in Eq.~(\ref{eq:dlam}) is expressed as: 
\begin{eqnarray}
  C & = & \frac{|<F|| \bm{A}||I>|^2}{2J_i+1}.
\end{eqnarray}
Here $|F>$ and $|I>$ represent the final and initial nuclear states, respectively, and 
$\bm{A}$ is the spatial component of the nuclear axial vector current. 

We emphasize that the energy differential decay rate, given in Eq.~(\ref{eq:dlam}), is
the product of the neutrino phase space factor and the atomic spectral function, which 
are independent of each other. 


We examine the role of excited states in the daughter atom (Tb) in the spectral function $P(E_{\rm ex})$.
Within a simple particle-hole picture, the spectral function, as defined in Eq.~(\ref{eq:spect}) 
is expressed in the standard form ~\cite{Bambynek:1977zz},
\begin{eqnarray}
  P(E_{\rm ex}) & = & \sum_x P_x(E_{\rm ex}),
\end{eqnarray}
with
\begin{eqnarray}
  P_x(E_{\rm ex}) & = & 2 n_x B_x \frac{\beta_x^2}{4\pi}
   \frac{\Gamma_x/(2\pi)}{(E_{\rm ex} - \epsilon_x)^2 + \Gamma_x^2/4}. \label{eq:p}
\end{eqnarray}
Here $x$ represents hole quantum number $x = (n,l,j)$.  
The energies of the excited states are approximated by the binding energy of the hole state $\epsilon_x$
and the width $\Gamma_x$. The term $\beta_x^2/(4\pi)$ represents the probability of the electron at the origin.
The parameters $n_x$ and $B_x$ represent the relative occupation number of the shell and
 the effects of electron exchange and overlap, respectively.  

 We considered the $s_{1/2}$ and $p_{1/2}$ states of $K, L, M, N, O$, and $P$ orbits, 
which have non-zero overlap with the nuclear transition density.  Among these, the
$M$ and $N$ orbits are of particular importance for the $^{159}$Dy EC.  
The parameters for the $M$ and $N$ states used in our estimation are given in Table~\ref{Tab-1}.
 The parameters $\epsilon_x$, $\Gamma_x$, $B_x$ and $\beta_x$ are sourced from
references ~\cite{Ge:2021gnp,thompson2009x,CAMPBELL:2001fbu,Bambynek:1977zz,band1986electron}.

\begin{center}
\begin{table}[h]
\begin{tabular}{c|cc|cc}\hline
     $x$        &  $M_1$ & $M_2$ & $N_1$  &  $N_2$ \\  \hline
$\epsilon_x$(keV) &\  1.968 &\  1.768 &\  0.396 & \ 0.322 \\
  $\Gamma_x$(eV) &\  13 &\  5.8 & \ 5.1 & \ 5.26 \\
  $\beta_x$  &\  0.212 &\  0.048 & \ 0.102 &\  0.023\\ \hline
\end{tabular}
\caption{Parameters of electron hole states.
The parameters of $\epsilon_x$ and $\Gamma_x$ correspond to Tb, while  $\beta_x$ pertains to Dy.}
\label{Tab-1}
\end{table}
\end{center}

The ROI for neutrino energy is $E_\nu \sim m_\nu$ ($E_{\rm ex} \sim Q - \Delta_{\rm nucl} - m_\nu $).
Hole states close to this region include 
$N$ states, referred to as   continuum states,  and $M$ states which are
  subthreshold states.
Their binding energies are 
approximately equidistant, about $0.78$ keV, from $Q - \Delta_{\rm nucl}$. 
The subthreshold states may influence the neutrino
spectrum due to their width.

\begin{center}
  \begin{figure}
    \includegraphics[width=10cm]{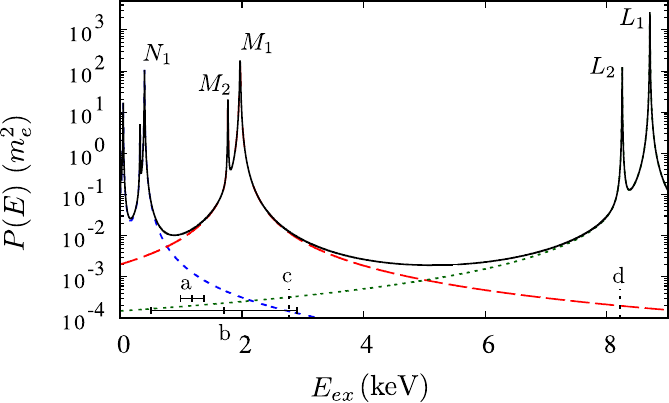}
    \caption{Spectral function $P(E_{\rm ex})$ of EC for $^{159}$Dy.
The solid black curve includes contributions of all hole states.
The short-dashed blue curve shows the contribution of the 
 continuum states. The dashed red curve and dotted green curve illustrate the 
contributions of the $M$ and $L$ states, respectively.
The symbol 'a' and 'b' mark positions with error of $E_{\rm ex}$ at $1.18(19)$ keV and $1.7(12)$ keV, respectively.  
Symbols 'c' and 'd' indicate positions of $E_{\rm ex}$ at $\epsilon_{M1} + 0.8{\rm keV}$ and $\epsilon_{L2} - 0.04{\rm keV}$,
respectively.}
    \label{fig-1}
  \end{figure}
\end{center}

Fig.~\ref{fig-1} illustrates the spectral function of EC for $^{159}$Dy. 
The solid black curve represents contributions from all hole states. 
The short-dashed blue curve shows the spectral function of the   continuum states, 
while the dashed red curve and dotted green curve depict the contributions of the $M$ and $L$ states, 
respectively, which are  subthreshold states. 
The symbol 'a' marks the excitation energy $E_{\rm ex}$ of $1.18(19)$ keV~\cite{Ge:2021gnp},
which is the primary focus of this study.  The symbol 'b' represents a previous measurement of 
the $Q$ value, determined to be $1.7(12)$ keV~\cite{huang2021ame,wang2021ame},
which is slightly older but still relevant for comparison.
At these excitation energies, the   subthreshold states, particularly $M$, primarily determine 
$P(E_{\rm ex})$, while the   continuum states contribute minimally.  
We observe enhancements by factors of $1.13 \times 10^1$ for 'a' and $3.02 \times 10^2$ for 'b', respectively,   
with the larger $Q$ value for 'b' leading to a greater enhancement.  
Historically, only   continuum states have been considered, 
but the significant increase in the rate due to the  subthreshold states makes the 
study of $^{159}$Dy particularly compelling for both $Q$ values of 'a' and 'b'. 
For the EC of $^{163}$Ho, $P(E_{\rm ex})$ near the neutrino at rest ($E_\nu \sim m_\nu$), 
marked by symbol 'c' in Fig.~\ref{fig-1}, originates dominantly from the  continuum states, 
associated with $M$ and only negligibly from the subthreshold states, associated with $L$.


More sophisticated theoretical studies of the spectral function have been
conducted ~\cite{brass2020ab}.
The estimation of the spectral function using Eq.~(\ref{eq:p}) requires improvements, 
particularly regarding the resonance energies estimated from the hole energy of the daughter 
atom and the energy-independent parameterization of the width. 
These improvements are crucial when the neutrino endpoint energy is close to the hole energy.

In the case of $^{159}$Dy EC, the energy region of interest lies between the two hole states,
and the yield depends relatively weakly on the excitation energy.  Furthermore, the neutrino 
spectrum of EC for Ho, where the charge of the daughter atom is
$Z=66$, has been measured~\cite{Velte:2019jvx}.  These measurements provide an excellent 
test of the spectral function of $^{159}$Dy after accounting for the neutrino phase space factor.
The data show that the shape of the spectral function around
$Q - \Delta_{\rm nucl} \sim 0$  is indeed flat.

\begin{center}
\begin{figure}[h]
  \includegraphics[width=15cm]{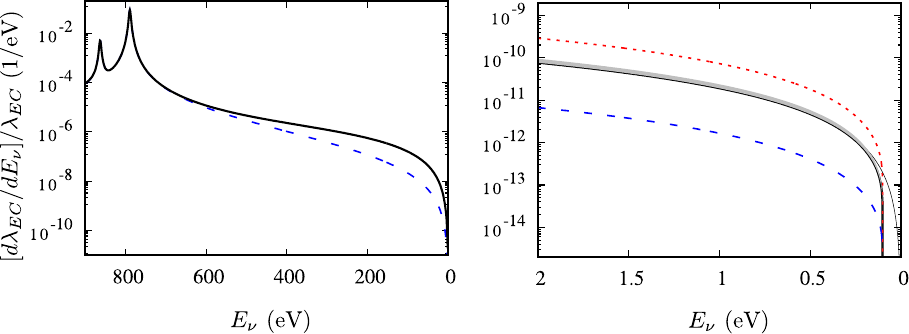}
  \caption{Normalized energy spectrum of neutrino.
The solid curve in the left panel includes contributions from all orbits, 
while the blue dashed line represents only the  continuum  $N$, $P$, and $O$ orbits. 
The right panel illustrates the neutrino energy spectrum near the endpoint. 
The black solid and blue dashed curves in the right panel correspond to those in the left panel, 
calculated with $m_\nu = 0.1$ eV, while the thin black solid curve represents $m_\nu = 0$ eV.
The uncertainty in the energy spectrum due to ($Q - \Delta_{\rm nucl} = 1.18 \pm 0.19\  \text{keV}$) is 
represented by the gray region.
The red dotted curve is derived from the solid curve by replacing the  $Q$ value
with $Q - \Delta_{\rm nucl} = 1.7$ keV.}
  \label{fig-2}
\end{figure}
\end{center}

Multiplying the spectral function by the neutrino phase space produces the experimental energy spectrum. 
The enhancement due to the inclusion of the $M$ orbit clearly reflects the distribution of the released energy. 
The normalized distribution of the released energy for $^{159}$Dy is shown in Fig.~\ref{fig-2}.
The solid curve in the left panel represents the results including contributions from all orbits, 
while the blue dashed line includes only the   continuum  $N$, $P$, and $O$ orbits. 
The contribution of the   subthreshold  $M$ orbit significantly enhances the rate, 
particularly in the region where the neutrino mass effect is most pronounced. 
The  neutrino energy distribution close to the minimum neutrino energy is depicted in the right panel. 
The black solid and blue dashed curves are the same as those in the left panel, calculated 
with $m_\nu = 0.1$ eV, while the thin black solid curve corresponds to $m_\nu = 0$ eV.
The uncertainty in the energy spectrum due to ($Q - \Delta_{\rm nucl} = 1.18 \pm 0.19\  \text{keV}$) is 
represented by the gray region.
Inclusion of the $M$ orbit enhances the partial rate by more than an order of magnitude in this region.
The red dotted curve is derived from the solid curve by simply replacing the $Q$ value
with $Q - \Delta_{\rm nucl} = 1.7$ keV.  The rate further increases because the larger $Q$ value 
shifts the endpoint closer to the  subthreshold $M$ states, supporting the feasibility 
of experimental exploration.  Since the decay rate in the ROI depends only weakly on the $Q$ value, 
an experiment can be designed without being constrained by variations in its measurement.

To further investigate the role of  subthreshold states, we define the ratio $r$.
\begin{eqnarray}
  r & = & \frac{\delta \lambda_{\rm EC}}{\lambda_{\rm EC}},
\end{eqnarray}
where $\delta \lambda_{\rm EC}$ represents the contribution of the endpoint region 
of the neutrino spectrum, $m_\nu < E_\nu < m_\nu + \delta$  to the EC rate. This is given by
\begin{eqnarray}
  \delta \lambda_{\rm EC} & = & \int_{m_\nu}^{m_\nu+\delta} \frac{d\lambda_{\rm EC}}{dE_\nu}dE_\nu.
\end{eqnarray}

Table~\ref{tab-1} shows the $r$ values for $^3$H beta decay, and the EC of $^{159}$Dy and $^{163}$Ho.
By including the   subthreshold states, $r$ increases by approximately one order of magnitude, 
which is larger than for $^3$H beta decay and the EC of $^{163}$Ho.
In the case of  $^{163}$Ho, the $K$ states are so far from the $Q$ value that the inclusion of 
  subthreshold states has little effect on $r$.  This is evident from the excitation function 
near 'c' 
in Fig.~\ref{fig-1}.
The excitation energy $E_{\rm ex}$ corresponding to  $E_\nu \sim 0$ for the EC of $^{163}$Ho, 
indicated as 'c' in Fig.~\ref{fig-1} is given as $E_{\rm ex} \sim \epsilon_{M1}^{Z=66}+0.8$ keV, assuming 
$Q \sim 2.8$  keV~\cite{PhysRevLett.119.122501,PENTATRAP:2024aed}.
Although the atomic number $Z$ differs, the qualitative features of $P(E_{\rm ex})$ are predominantly 
determined by the relative position between the hole state energy and the $Q$ value.

\begin{table}
  \begin{tabular}{c|c|c|c}\hline
    $r$ & $^3$H $\beta$ decay & EC $^{159}$Dy & EC $^{163}$Ho \\ \hline
    $m_\nu=1$ eV &   $1.5\times 10^{-12}$ &  $3.3\times 10^{-11}$($2.9\times 10^{-12}$)
    &  $2.6\times 10^{-12}$($2.5\times 10^{-12}$)\\
    $m_\nu=0.1$ eV &  $3.8\times 10^{-13}$ & $8.4\times 10^{-12}$($7.3\times 10^{-13}$)
    &  $6.5\times 10^{-13}$($6.2\times 10^{-13}$)\\ \hline
  \end{tabular}
  \caption{Values of $r$ for $^3$H beta decay and EC of $^{159}$Dy and $^{163}$Ho for 
$m_\nu = 0.1$ and $1$ eV and $\delta = 1$ eV.
    For EC, both   continuum states and   subthreshold states are included, 
with the numbers in parentheses showing results excluding contribution from   subthreshold states.}
  \label{tab-1}
  \end{table}

Using the calculated $r$, when comparing the decay rates of the ROI 
($r \times \text{branching ratio}/\text{lifetime}$), the decay rate of $^{159}$Dy, which depends on the $Q$ value, 
is 2-3 orders of magnitude smaller than that of that of $^3$H. $^3$H currently serves as the standard 
in neutrino mass experiments and has shown remarkable success, most notably demonstrated by the KATRIN 
experiment ~\cite{katrin2022direct}.  Despite its lower decay rate, $^{159}$Dy holds promise for future 
applications and experimental advancements for the following reasons.  

As a metal, Dy has a density significantly higher than that of a gas. Recent breakthroughs in 
superconducting devices, such as TES (transition-edge sensors), have demonstrated exceptional 
capabilities, achieving an energy resolution of 0.3 eV for 0.8 keV X-rays ~\cite{sakai2020demonstration}.  
With nearly 100\% detection efficiency, TES-based methodologies combined with segmentation techniques 
offer a compelling solution to overcome the issues associated with the low decay rate of $^{159}$Dy. 
As measurement sensitivity continues to improve, the spectrum near the endpoint becomes 
increasingly critical. In this study, $P(E)$ is an independent parameter that can be experimentally determined. 
In addition to Ho, used as a reference in our calculations, one can measure $P(E)$ in $^{159}$Dy 
by using states with different $Q$ values, which facilitates the independent extraction of $P(E)$ in the future 
experiments.

Another interesting case is  $^{111}$In~\cite{ge2022high}.
The effective $Q$ value for the $7/2^+$ excited state of $^{111}$Cd 
is $Q - \Delta_{\rm nucl} \sim 3.69(19)$ keV, which is very close to the  subthreshold   hole state of L2,
$\epsilon_{L2} \sim 3.727$ keV.  We would like to highlight that even if $L2$ lies below the endpoint energy
(subthreshold state),  $r$ is substantially enhanced by approximately $10^4$ when including 
the subthreshold state, making the $^{111}$In EC particularly intriguing.  
This phenomenon can be understood from the excitation function around 'd' in Fig.~\ref{fig-1},
which indicates $E_{\rm ex} = \epsilon_{L2}^{Z=65} - 0.04$ keV, corresponding to the central value of the experiment.


In summary, we have examined the final residual energy distribution in the electron capture process of 
$^{159}$Dy.  The atomic system's response is characterized by the spectral function $P(E_{\rm ex})$, 
which remains independent of reaction kinematics.  As a result, both the continuum state and 
 subthreshold state  must be accounted for in all EC processes.  
The inclusion of 'virtual EC' for the  subthreshold state significantly enhances the rate 
increasing it by more than an order of magnitude in the region of zero-momentum neutrino emission 
for $^{159}$Dy.   Conversely, no adjustments are required for previous findings concerning $^{163}$Ho.

\medskip

This work was supported by JSPS KAKENHI Grant Number JP22H04946 and JP24H00225. 
The author (TK) expresses gratitude to Professor K. Ichimura for his insightful discussions on TES.

\bibliography{dy159ec.bib}

\end{document}